\documentclass[manuscript]{acmart}

\acmYear{2019}
\setcopyright{rightsretained}

\acmConference[PyCon '19]{Israeli Python Conference 2019}{May 2019}{Israel}
\acmDOI{10.13140/RG.2.2.36809.77925/1}

\AtBeginDocument{%
  \providecommand\BibTeX{{%
    \normalfont B\kern-0.5em{\scshape i\kern-0.25em b}\kern-0.8em\TeX}}}

\usepackage{float}
\usepackage{soul}
\usepackage{color}
\usepackage{bbm}
\usepackage{enumitem}
\usepackage{listings}

\definecolor{Blue}{rgb}{0,0,1}

\definecolor{Orange}{rgb}{1,0.5,0}

\definecolor{Green}{rgb}{0,1,0}

\begin{document}

\title{Social Network Analysis: From Graph Theory to Applications with Python}

\author{Dmitri Goldenberg}
\email{dima.goldenberg@booking.com}
\affiliation{\institution{Booking.com, Tel Aviv}}

\begin{abstract}


Social network analysis is the process of investigating social structures through the use of networks and graph theory. It combines a variety of techniques for analyzing the structure of social networks as well as theories that aim at explaining the underlying dynamics and patterns observed in these structures. It is an inherently interdisciplinary field which originally emerged from the fields of social psychology, statistics and graph theory. This talk will covers the theory of social network analysis, with a short introduction to graph theory and information spread. Then we will deep dive into Python code with NetworkX to get a better understanding of the network components, followed-up by constructing and implying social networks from real Pandas and textual datasets. Finally we will go over code examples of practical use-cases such as visualization with matplotlib, social-centrality analysis and influence maximization for information spread.

\end{abstract}

\ccsdesc[500]{Human-centered computing~Social network analysis}

\keywords{Social Network Analysis, Python}

\maketitle

\section{Introduction}\label{sec:intro}
Social network analysis is the process of investigating social structures through the use of networks and graph theory. This article introduces data scientists to the theory of social networks, with a short introduction to graph theory, information spread and influence maximization \cite{goldenberg2018timing}. It dives into Python code with NetworkX \cite{SciPyProceedings_11} constructing and implying social networks from real datasets. A video version of this article is available on \textit{Pycon Youtube channel}\footnote{\url{https://www.youtube.com/watch?v=px7ff2_Jeqw}}.

\begin{figure}[]
    \centering
    \includegraphics[width=1\textwidth]{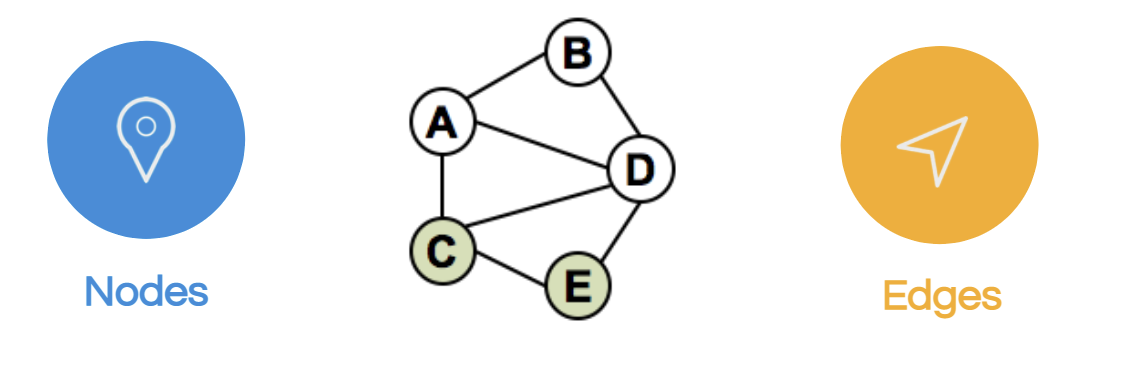}
    \caption{Example network and components}
    \label{fig:demo}
\end{figure}

\section{Network Theory}

\subsection{Network Components}
We’ll start with a brief intro in network’s basic components: nodes and edges.

\textbf{Nodes} (A,B,C,D,E in the example) are usually representing entities in the network, and can hold self-properties (such as weight, size, position and any other attribute) and network-based properties (such as Degree- number of neighbours or Cluster- a connected component the node belongs to etc.).

\textbf{Edges} represent the connections between the nodes, and might hold properties as well (such as weight representing the strength of the connection, direction in case of asymmetric relation or time if applicable).

These two basic elements can describe multiple phenomena, such as social connections, virtual routing network, physical electricity networks, roads network, biology relations network and many other relationships.

\subsection{Real-world networks}
\begin{figure}[b]
    \centering
    \includegraphics[width=1\textwidth]{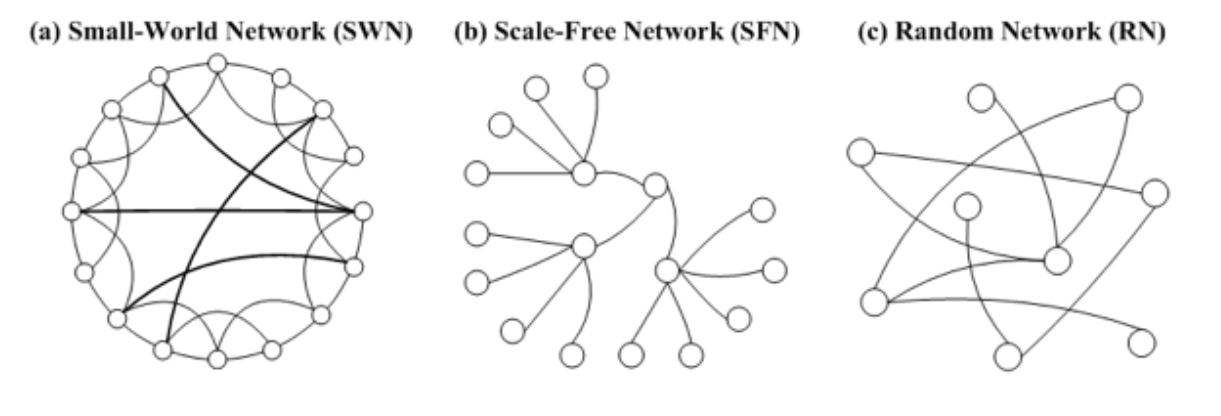}
    \caption{Complex networks (taken from \cite{huang2005influence}) }
    \label{fig:complex}
\end{figure}

Real-world networks and in particular social networks have a unique structure which often differs them from random mathematical networks. Figure \ref{fig:complex} provides examples of complex networks (taken from \cite{huang2005influence}).

\begin{itemize}
    \item \textbf{Small World} phenomenon \cite{kleinberg2000small} claims that real networks often have very short paths (in terms of number of hops) between any connected network members. This applies for real and virtual social networks (the six handshakes theory) and for physical networks such as airports or electricity of web-traffic routing.
    \item \textbf{Scale Free} \cite{barabasi2000scale} networks with power-law degree distribution have a skewed population with a few highly-connected nodes (such as social-influences) and a lot of loosely-connected nodes.
    \item \textbf{Homophily} \cite{mcpherson2001birds} is the tendency of individuals to associate and bond with similar others, which results in similar properties among neighbors.
\end{itemize}

\subsection{Centrality Measures}
\begin{figure}
    \centering
    \includegraphics[width=1\textwidth]{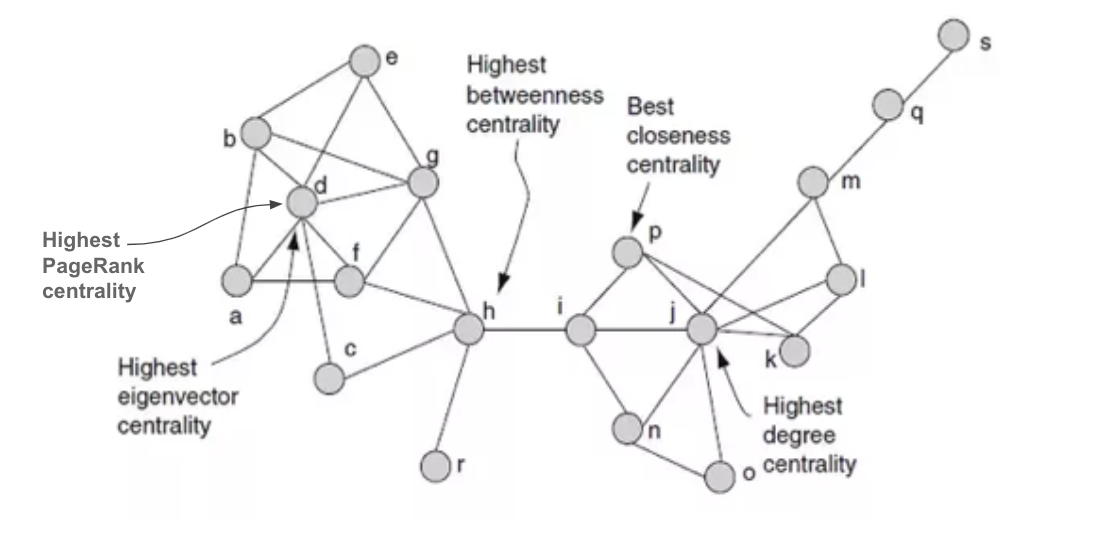}
    \caption{Illustration of various centrality measures (taken from \cite{ortiz2010discovering}).}
    \label{fig:central}
\end{figure}

Highly central nodes play a key role of a network, serving as hubs for different network dynamics. However the definition and importance of centrality might differ from case to case, and may refer to different centrality measures, as depicted in figure \ref{fig:central} (taken from \cite{ortiz2010discovering}).

\begin{itemize}
    \item \textbf{Degree} — the amount of neighbors of the node
    \item \textbf{EigenVector} \cite{bonacich2007some} /  \textbf{PageRank} \cite{page1999pagerank} — iterative circles of neighbors
    \item \textbf{Closeness} \cite{okamoto2008ranking} — the level of closeness to all of the nodes
    \item \textbf{Betweenness} \cite{brandes2001faster} — the amount of short path going through the node

\end{itemize}

Different measures can be useful in different scenarios such web-ranking (page-rank), critical points detection (betweenness), transportation hubs (closeness) and other applications.

\section{Building a Network}

\begin{figure}[b]
    \centering
    \includegraphics[width=1\textwidth]{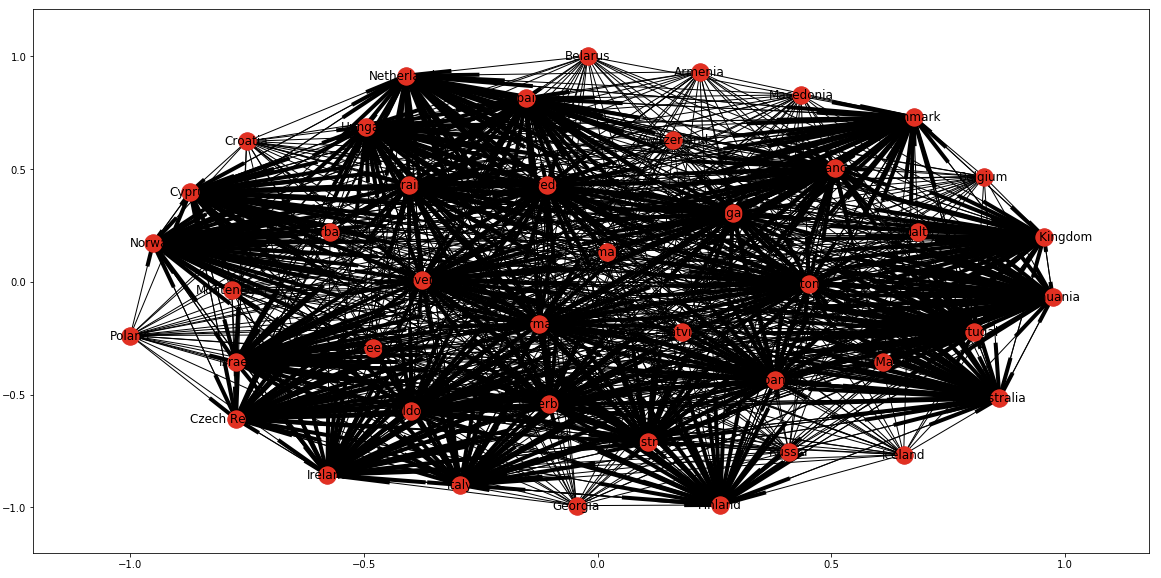}
    \caption{nx.draw\_networkx(G) outcome on Eurovision 2018 votes network
}
    \label{fig:mess}
\end{figure}

Networks can be constructed from various datasets, as long as we’re able to describe the relations between nodes. 
In the following example we’ll build and visualize the Eurovision 2018 votes network 
(based on official data\footnote{\url{https://eurovision.tv/story/the-results-eurovision-2018-dive-into-numbers}}) with Python networkx \cite{SciPyProceedings_11} package.
We’ll read the data from excel file to a pandas dataframe to get a tabular representation of the votes. Since each row represents all of the votes of each country, we’ll melt the dataset to make sure that each row represents a single vote (edge) between two countries (nodes).
Then, we will build a directed graph using networkx from the edgelist we have as a pandas dataframe. Finally, we’ll try the generic method to visualize, as shown in code \ref{code:build} (full code can be found at Github repository\footnote{\url{https://github.com/dimgold/pycon_social_networkx}})

\begin{lstlisting}[caption={Creating a network with Networkx},label={code:build},language=Python]
votes_data = pd.read_excel('data.xlsx')
votes_melted = votes_data.melt(
    ['Rank','Running order', 'Country','Total'],
    var_name = 'Source Country',value_name='points') 
    
G = nx.from_pandas_edgelist(votes_melted, source='Source Country',
            target='Country', edge_attr='points',   create_using=nx.DiGraph())
nx.draw_networkx(G)
\end{lstlisting}

\subsection{Visualization}
Unfortunately the built-in draw method results in a very incomprehensible figure as shown in figure \ref{fig:mess}. The method tries to plot a highly connected graph, but with no useful “hints” it’s unable to make a lot of sense from the data. We will enhance the figure by dividing and conquering different visual aspects of the plot with a prior knowledge that we have about the entities:
\begin{itemize}
    \item \textbf{Position} — each country is assigned according to its geo-position
    \item \textbf{Style} — each country is recognized by its flag and flag colors
    \item \textbf{Size} — the size of nodes and edges represents the amount of points
\end{itemize}

Plotting of the network components in parts is shown in code \ref{code:plot}.

\begin{figure}[b]
    \centering
    \includegraphics[width=1\textwidth]{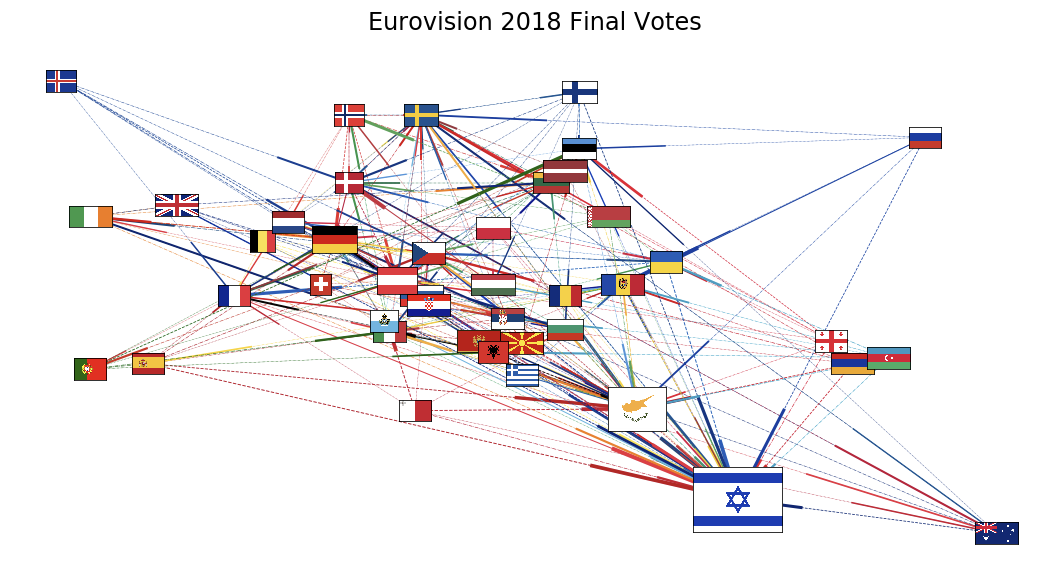}
    \caption{Step-by-step Eurovision network plot}
    \label{fig:clean}
\end{figure}

\begin{lstlisting}[caption={Network components plotting},label={code:plot},language=Python]
for e in G.edges(data=True):
    width = e[2]['points']/24 
    style=styles[int(width*3)]
    nx.draw_networkx_edges(G,pos, edgelist=[e],width=width, style=style,
        edge_color = RGB(*flag_color[e[0]] )
for node in G.nodes():      
    imsize = G.in_degree(node,  weight='points')
    flag = mpl.image.imread(flags[node])
    (x,y) = pos[node]
    xx,yy = trans((x,y))
    xa,ya = trans2((xx,yy))
    country = plt.axes(   [xa-imsize/2.0,  ya-imsize/2.0,  imsize, imsize ])
    country.imshow(flag)
    country.set_aspect('equal')
    country.tick_params(**tick_params)
\end{lstlisting}

The new figure \ref{fig:clean} is a bit more readable, and giving us a brief overview of the votes. As a general side-note, plotting networks is often hard and requires to perform thoughtful tradeoffs between the amount of data presented and the communicated message. (You can try to explore other network visualization tools such as Gephi , Pyvis or GraphChi).

\section{Information Flow}

Information diffusion \cite{shakarian2015diffusion} process may resemble a viral spread of a disease, following contagious dynamics of hopping from one individual to his social neighbors. Two popular basic models are often used to describe the process:

\textbf{Linear Threshold} \cite{shakarian2013scalable} defines a threshold-based behavior, where the influence accumulates from multiple neighbors of the node, which becomes activated only if the cumulative influence passed a certain threshold according to the following formula:
$$ \sum_{active_u} W_{uv} \geq \theta_v $$
Such behavior is typical to movie recommendations, where a tip from of one of your friends might eventually convince you to see a movie, after hearing a lot about it.

In the \textbf{Independent Cascade} model \cite{goldenberg2001talk}, each of the node’s active neighbors has a probabilistic and independent chance to activate the node. This resembles a viral virus spread, such as in Covid-19, where each of the social interactions might trigger the infection.

\subsection{Information Flow Example}
To illustrate an information diffusion process we’ll use the Storm of Swords network, based on Game of Thrones show characters. The network was constructed based on co-appearance in the “Song of Ice and Fire books” \cite{beveridge2016network}.

Relying on the independent cascade model, we’ll try to track down rumor spreading dynamics, which are quite common in this show.
Suppose \textit{Jon Snow} knows nothing at the beginning of the process, while his two loyal friends, \textit{Bran Stark} and \textit{Samwell Tarly}, know a very important secret about his life. Let’s watch how the rumor spreads under the Independent Cascade model:

\begin{lstlisting}[caption={Independent Cascade Code},label={code:ind_cas},language=Python]
def ind_cascade(G,t,infection_times):
    max_w = max([e[2]['weight'] for e in G.edges(data=True)])
    current_infectious = [n for n in infection_times 
                         if infection_times[n]==t]
    for n in current_infectious:
        for v in G.neighbors(n):
            if v not in infection_times:
                if G.get_edge_data(n,v)['weight']>=np.random.random()*max_w:
                    infection_times[v] = t+1
    return infection_times

inf_times = {'Bran-Stark':-1, 'Samwell-Tarly':-1,'Jon-Snow':0}
for t in range(10):
    plot_G(subG,pos,infection_times,t)
    inf_times=ind_cascade(subG,t,inf_times)
\end{lstlisting}

\begin{figure}[b]
    \centering
    \includegraphics[width=1\textwidth]{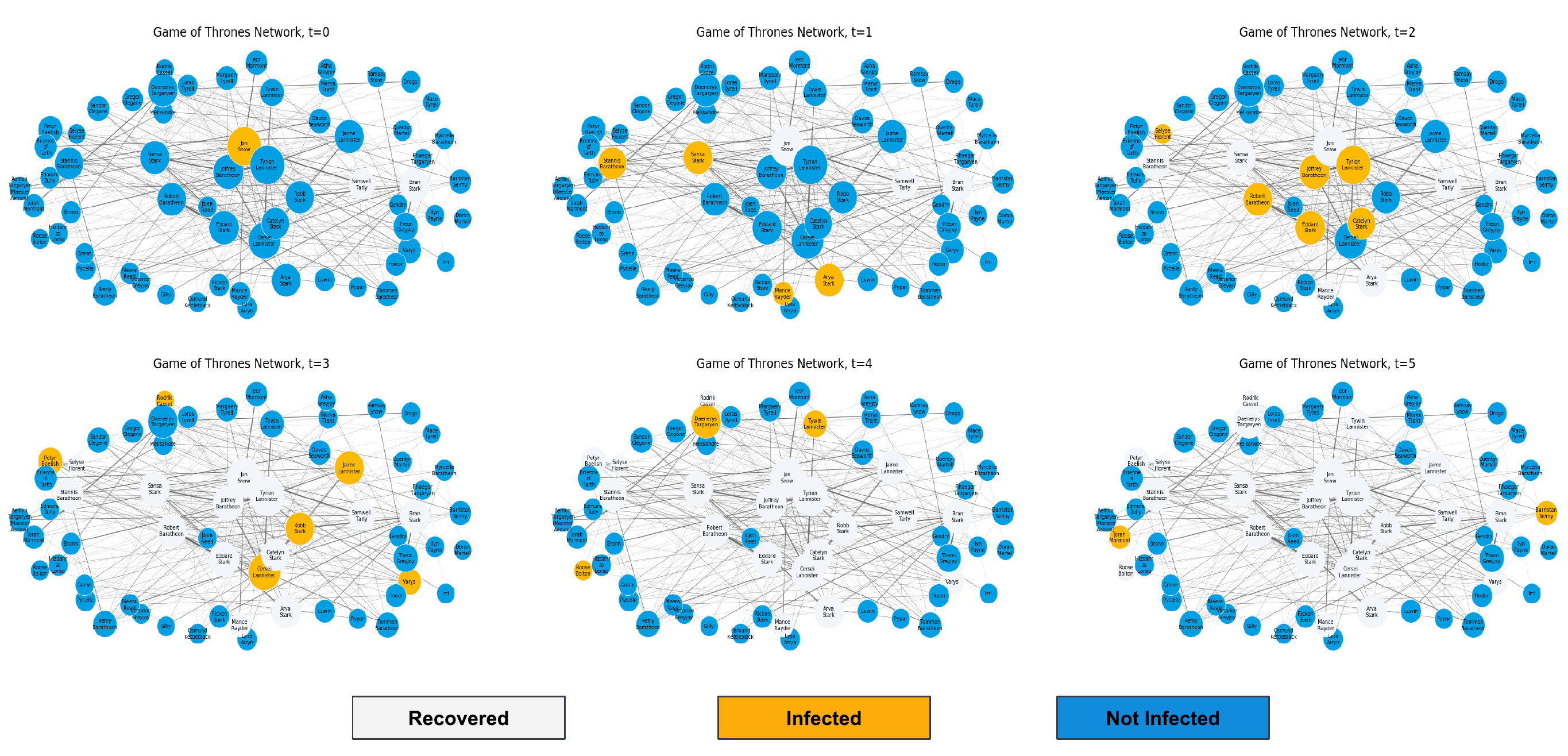}
    \caption{Independent Cascade diffusion simulation on Game of Thrones network}
    \label{fig:got}
\end{figure}

As shown in figure \ref{fig:got} the rumor reaches Jon at t=1, spreads to his neighbors in the following time-steps and quickly spreads all around the network, resulting in being a public knowledge.
Such dynamics are highly dependent on model parameters, which can drive the diffusion process to different patterns.

\section{Influence Maximization}

The influence maximization problem \cite{kempe2003maximizing} describes a marketing (but not only) setup, where the goal of the marketer is to select a limited set of nodes in the network (seeding set) such that will naturally spread the influence to as much nodes as possible. For example, consider inviting a limited amount of influencers to a prestigious product launch event, in order to spread the word to the rest of their network.
Such influencers can be identified with numerous techniques, such as using the centrality measures \cite{borgatti2005centrality, hinz2011seeding} we’ve mentioned above. The most central nodes in Game of Thrones network according to different measures are listed in table \ref{tab:center}.

\begin{table}[]

\caption{Leading characters per centrality measure}
\label{tab:center}
\begin{tabular}{|l|l|l|l|l|l|l|l|}

\hline
\multicolumn{2}{|l|}{\textbf{Degree}} & \multicolumn{2}{l|}{\textbf{Weighted Degree}} & \multicolumn{2}{l|}{\textbf{Pagerank}} & \multicolumn{2}{l|}{\textbf{Betweenness}} \\ \hline
Name               & Score            & Name                   & Score                & Name                & Score            & Name                 & Score              \\ \hline
Tyrion             & 40               & Tyrion                 & 1842                 & Tyrion              & 0.052            & Robert               & 0.22               \\ \hline
Robert             & 37               & Cersei                 & 1627                 & Jon Snow            & 0.048            & Brienne              & 0.12               \\ \hline
Joffrey            & 35               & Joffrey                & 1518                 & Cersei              & 0.046            & Rodrik               & 0.11               \\ \hline
\end{tabular}
\end{table}

As we can see in table \ref{tab:center}, some of the characters re-occur at the top of different measures, and are also well known for their social influence in the show.
By simulating the selection of most central nodes we observe that picking a single node of the network can achieve about 50\% of network coverage — That’s how important social influencers might be.

\begin{figure}
    \centering
    \includegraphics[width=1\textwidth]{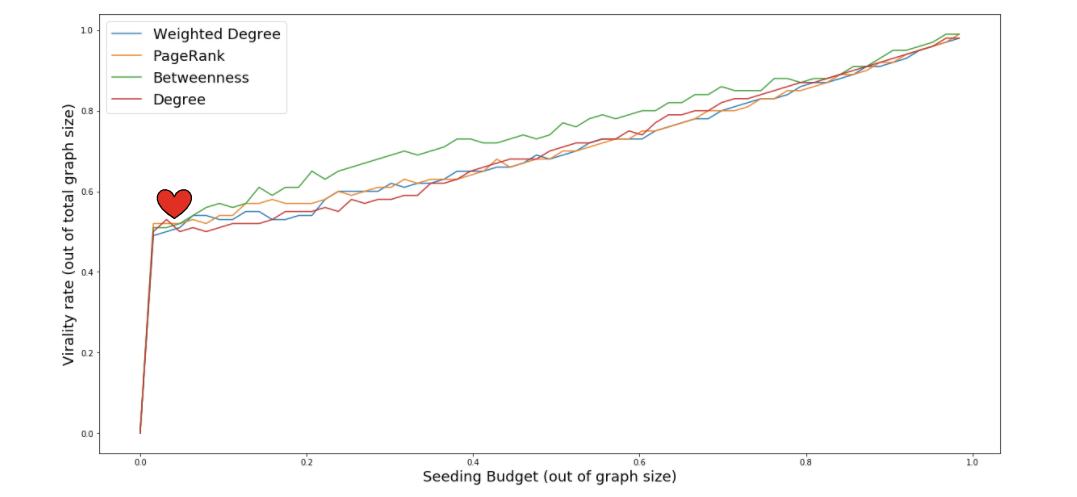}
    \caption{Network influence coverage by different methods and budgets}
    \label{fig:res}
\end{figure}

On the other hand Influence Maximization is Hard. In fact it’s considered as an NP-Hard problem. Many heuristics were developed to find the best seeding set in an efficient calculation. Trying a brute-force method to find the best seeding couple in our network resulted in spending 41 minutes and achieving 56\% of coverage (by selecting Robert Baratheon and Khal Drogo)- a result that would be hard to achieve with centrality heuristics.

\section{Conclusion}

Network analysis is a complex and useful tool for various domains, in particular in the rapidly growing social networks. The applications of such analysis include marketing influence maximization, fraud detection or recommender systems. There are multiple tools and techniques that can be applied on network datasets, but they need to be chosen wisely, taking into account the problem’s and the network’s unique properties.

\bibliographystyle{ACM-Reference-Format}
\bibliography{sample-base}

\end{document}